# 600-T Magnetic Fields due to Cold Electron Flow in a simple Cu-Coil irradiated by High Power Laser pulses

Baojun Zhu[1], Yutong Li[1,7*], Dawei Yuan[2], Yanfei Li[1], Fang Li[1], Guoqian Liao[1], Jiarui Zhao[1], Jiayong Zhong[3], Feibiao Xue[4], Huigang Wei[2], Kai Zhang[2], Bo Han[2], Xiaoxing Pei[2], Chang Liu[3], Zhe Zhang[1], Weimin Wang[1], Jianqiang Zhu[5], Gang Zhao[2] and Jie Zhang[6,7]

[1] National Laboratory for Condensed Matter Physics, Institute of Physics, Chinese Academy of Sciences, Beijing 100190, China

[2] National Astronomical Observatories, Chinese Academy of Sciences, Beijing 100012, China

[3] Department of Astronomy, Beijing Normal University, Beijing 100875, China

[4] Institute of Nuclear Physics and Chemistry, CAEP, P.O. Box 919-212, Mianyang 621900, China

[5] National Laboratory on High Power Lasers and Physics, Shanghai, 201800, China

[6] Key Laboratory for Laser Plasmas (MoE) and Department of Physics, Shanghai Jiao Tong University, Shanghai 200240, China

[7] IFSA Collaborative Innovation Center, Shanghai 200240, China

A new simple mechanism due to cold electron flow to produce strong magnetic field is proposed. A 600-T strong magnetic field is generated in the free space at the laser intensity of $5.7 \times 10^{15}\ W \cdot cm^{-2}$. Theoretical analysis indicates that the magnetic field strength is proportional to laser intensity. Such a strong magnetic field offers a new experimental test bed to study laser-plasma physics, in particular, fast-ignition laser fusion research and laboratory astrophysics.

Introduction

Laboratory generation of large, strong magnetic fields is of significance to many research areas including plasma and beam physics [1], astrophysics [2], material science [3], and atomic and molecular physics [4]. In material science, strong magnetic fields offer new opportunities to observe the magnetization process of the highly frustrated materials [5]. In astrophysics, soft γ-ray repeaters (SGRs) generate extreme surface magnetic fields of $\sim 10^{15}$ gauss [6-9]. Jet formation [10], magnetic reconnection [11, 12] and collisionless shock generation [1] due to strong magnetic fields are also important research topics attracting much interest. Laboratory generation of large, strong magnetic fields makes it possible to study them in controllable experiments [13-15].

---

* Electronic address: ytli@iphy.ac.cn

At present, traditional strong magnets including resistive magnets, superconducting magnets and hybrid magnets can produce stable magnetic field up to 45 T, repetitive pulses up to 100 T and a single-shot pulse up to 300 T [16]. However, the power consumption of resistive magnets is very high, and the high stress caused by high current and strong magnet is also a great challenge for the development of resistive magnets. Superconducting magnets are limited by the special critical current and critical magnetic field effect of superconducting materials. The applied superconducting magnets with about 50-milimeter aperture can produce about 20 T magnet field, and with the increase of the aperture, the magnet field strength produced by superconducting circle will decrease [17].

With the rapid development of high power laser technology, kilo-ampere [18] or even mega-ampere [19] current is generated by laser-plasma interactions. This offers a new opportunity to produce pulsed strong magnetic fields. Now, several generation mechanisms have been proposed in laser-produced plasma experiments. Spontaneous magnetic fields can be generated by nonparallel temperature and density gradients in ablated plasmas [20]. Surface-magnetic fields can be generated by oblique intense laser irradiation [21, 22]. Although the spontaneous magnetic fields and the surface-magnetic fields in plasma are very strong, they are not applicable for applications in free space. In 1986, H. Daido *et al.* demonstrated a strong magnetic field with a capacitor-coil target irradiated by high power $CO_2$ laser pulses [23]. The field is generated due to the potential difference across two parallel plates induced by hot electrons. The potential drives a circulation of a strong reverse current through the coil connecting the plates. In 2005, C. Courtois *et al.* indicated that 100-T magnetic field was accessible using a capacitor-coil target [24]. In 2013, Fujioka *et al.* demonstrated a magnetic field up to 1 kT produced by nanosecond laser irradiating capacitor-coil targets [19].

In this paper, we propose a different mechanism to produce strong magnetic fields. Laser pulses are focused onto a metal target. Escaping of the hot electrons into vacuum from the target will lead to a large potential at the laser focal spot. Then, the cold electrons around is driven to the focus by the target potential. This current of the cold background electrons can excite a magnetic field pulse in free space. Our experiments demonstrate that a strong magnetic field with a maximum of 600 T is generated at laser intensity of $5.7 \times 10^{15}\ W \cdot cm^{-2}$.

Experiment

The experiments were performed at the SG-II laser facility, which can deliver a total energy of 2.0 kJ in a nanosecond Gaussian shape pulse. Fig. 1 shows the experimental setup. The target consisted of two parts, a Cu coil and a Cu planar target. The coil was made of a 200-μm-diameter Cu wire. The diameter of the coil was 1.16-mm-diameter. The coil was connected with the planar target. The planar target was perpendicular to the x axis. The coil plane was parallel to x-y plane. Eight heater laser beams at a wavelength of 351 nm were divided into two bunches and focused onto the two sides of the planar target simultaneously. The focal spot was measured to be about 150 μm FWHM in diameter. During the laser irradiating the planar target, hot electrons were generated and escaped into vacuum from the plasma. A large electrostatic potential grew up near the laser focus and charged the target. The potential dragged the background cold electrons to the focus to neutralize the target. This cold electron current would create a strong magnetic field pulse at the coil. The magnetic field was measured by a B-dot, which consisted of two 1.05-mm-diameter, reversed polarity single induction coils. The B-dot was positioned 25-mm away from coil center and 53-degree to the y axis. This type of magnetic probe can mitigate the electrostatic pickup created by the plasma [25]. A pinhole camera with magnification of 5 was used to monitor the X-ray emission from the target.

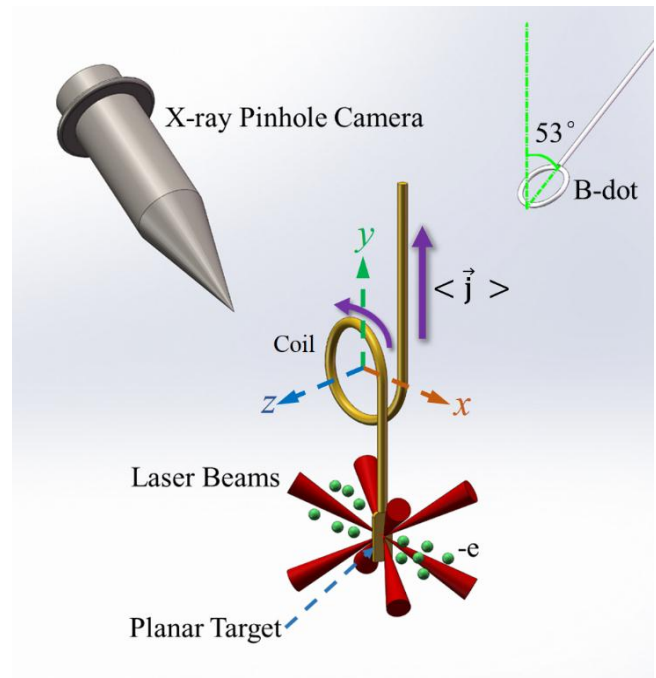

FIG. 1. Schematic view of the experimental setup at the SG-II laser facility showing the target, laser beams, B-dot and X-ray pinhole camera. Eight heater laser beams at a wavelength of 351 nm were divided into two bunches and focused onto the two sides of the planar target simultaneously. Hot electrons (green small balls) generated and escaped into vacuum from the target, resulting in a large electrostatic potential. A strong return current (purple arrow) due to cold electrons was generated. Then, a strong magnetic field was excited and detected by the B-dot (white coil). The X-ray pinhole camera was used to monitor the X-ray emission from the target.

Results and discussions

Fig. 2 shows the induced voltage signals measured by two reversed polarity single-turn induction coils at the laser intensity of $5.7 \times 10^{15}\ W \cdot cm^{-2}$. Since the electrostatic potential is identical on both coils, the final signal $V$ can be excited by subtracting out the induced common voltage [25]. The integration of $V_{mea}(t)$ is the magnetic field strength. The peak magnetic intensity, measured at the local B-dot position, is 0.031 T.

To distinguish the magnetic field generated by the coil with other possible sources, a straight Cu wire target without the coil was irradiated by the laser pulses. The magnetic field strength measured by the B-dot is only 0.006 T, much lower than 0.031 T. This small magnetic field is probably due to the spontaneous field generated by laser-plasma interactions. Therefore, the main source of the free-space magnetic field measured is from the coil target.

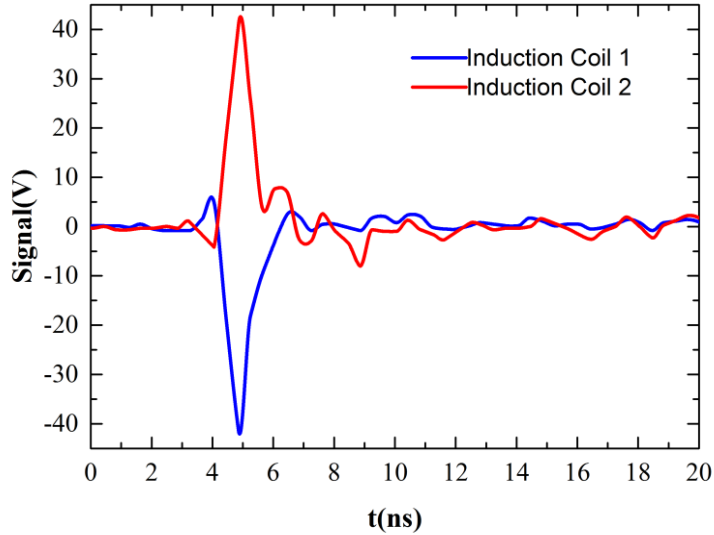

FIG. 2. The typical magnetic probe signals which corresponds to the time derivative of the magnetic field at the laser intensity of $5.7 \times 10^{15} W \cdot cm^{-2}$

By matching the measured magnetic field strength and that calculated with the initial shape of the coil target by Radia code [26], the current in the coil target is estimated to be 0.25 MA at the laser intensity of $5.7 \times 10^{15} W \cdot cm^{-2}$. The maximum magnetic field, $B_{max}$, is about 588 T, and the one at the coil center is about 257 T. Fig. 3 shows the 2-D magnetic field distribution of x-y plane with z=0 at the laser intensity of $5.7 \times 10^{15} W \cdot cm^{-2}$ calculated by Radia code. It demonstrates the field is mainly around the wire.

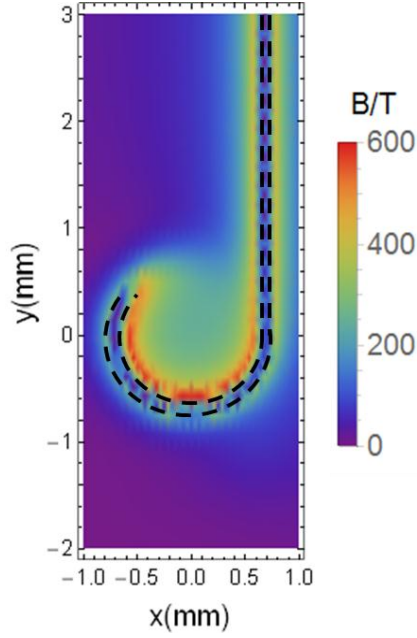

FIG. 3. The 2-D magnetic field distribution of x-y plane with z=0. The dashed line is the outline of the coil in the x-y plane. The field is mainly around the wire.

| Table 1. Summary of magnetic field strength, and current in the coil | | | | | |
|---|---|---|---|---|---|
| Energy | Laser Intensity | B@B-dot | Current | B@coil center | Maximum B |
| J | W cm-2 | T | MA | T | T |
| 536.24 | 1.52×1015 | 0.0065 | 0.054 | 55 | 126 |
| 1004.27 | 2.84×1015 | 0.011 | 0.088 | 90 | 206 |
| 1966.8 | 5.56×1015 | 0.031 | 0.25 | 257 | 588 |

The dependence of the magnetic field strength on the laser intensity is also investigated. Table 1 summarizes the details of the magnetic fields at different laser intensities. To see the dependence clearly, the variation of the $B_{max}$ and the B field at the coil center as a function of the laser intensity $I$ are also shown in Fig. 4. The field strength increases with the laser intensity $I$. The $B_{max}$ is obtained at the strongest laser intensity of $5.7 \times 10^{15} W \cdot cm^{-2}$. By fitting the data, we find that the $B_{max}$ is proportional to $I^{1.4\pm0.2}$.

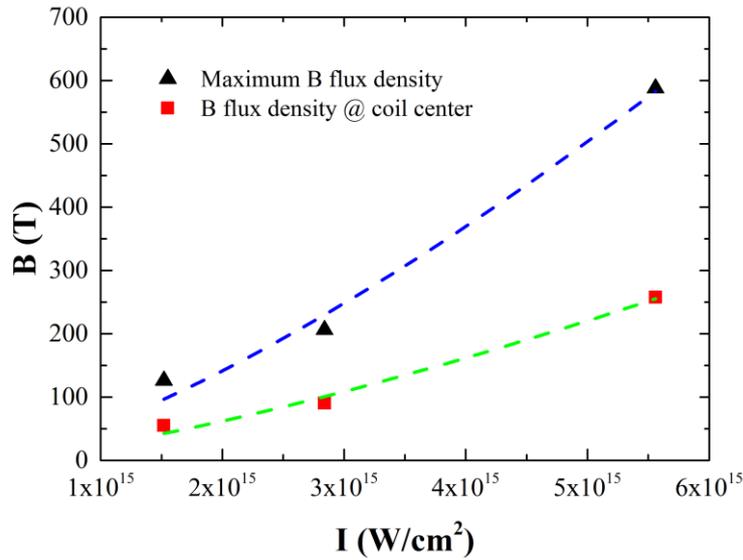

Fig. 4. Variation of the maximum magnetic field strength and the one at the coil center as a function of the laser intensity $I$. The blue and green dotted line both show a similar proportional relation as $B \propto I^{1.4\pm0.2}$.

## Conclusions

We have demonstrated a maximum 600-T strong magnetic field with a simple coil target. The field strength increases sharply with the laser intensity. Compared with the capacitor-coil target, the generation mechanism of the coil target is straightforward and it is easy to be fabricated. Such a strong magnetic field can be applied to many research areas, in particular, laboratory astrophysics.

## Acknowledgements

The authors thank the staff of SG-II laser facility for operating the laser and target area. This work is supported by National Basic Research Program of China (Grant No.2013CBA01501) and the National Natural Science Foundation of China (Grant No. 11135012 and 11375262).